\documentclass[aps,prb,reprint,amsfonts,amssymb,amsmath,superscriptaddress,longbibliography]{revtex4-2}

\pdfoutput=1

\usepackage{amssymb,amsmath,bm}
\usepackage{fullpage}
\usepackage{graphicx}
\usepackage{bibunits}
\usepackage{amsmath}
\usepackage{wrapfig}
\usepackage{floatflt}
\usepackage{color,soul}
\usepackage{xspace}
\usepackage{dsfont}
\usepackage[official,right]{eurosym}
\usepackage[top=0.8in, bottom=0.8in, left=0.75in, right=0.75in]{geometry}
\usepackage[utf8]{inputenc}
\usepackage[colorlinks=true,urlcolor=blue,citecolor=blue,linkcolor=blue,bookmarks=false]{hyperref}
\usepackage[flushleft]{threeparttable}
\usepackage{array,booktabs,makecell}

\newcommand{\NBWO}{Nd$_{3}$BWO$_9$\xspace}
\newcommand{\NBeWO}{Nd$_{3}\,^{11}$BWO$_9$\xspace}

\newcommand{\be}{\begin{equation}}
\newcommand{\ee}{\end{equation} }
\newcommand{\bea}{\begin{eqnarray} }
\newcommand{\eea}{\end{eqnarray} }

\begin{document}

\title{Magnetic phase diagram of the breathing-kagome antiferromagnet \NBWO}
\author{D.~Flavi{\'a}n}
\email{daniefla@ethz.ch}
\affiliation{Laboratory for Solid State Physics, ETH Z{\"u}rich, 8093 Z{\"u}rich, Switzerland}

\author{J.~Nagl}
\affiliation{Laboratory for Solid State Physics, ETH Z{\"u}rich, 8093 Z{\"u}rich, Switzerland}

\author{S.~Hayashida}
\affiliation{Laboratory for Solid State Physics, ETH Z{\"u}rich, 8093 Z{\"u}rich, Switzerland}
\affiliation{Max-Planck-Institut f\"ur Festk\"orperforschung, Heisenbergstraße 1, 70569 Stuttgart, Germany}

\author{M.~Yan}
\affiliation{Laboratory for Solid State Physics, ETH Z{\"u}rich, 8093 Z{\"u}rich, Switzerland}

\author{O.~Zaharko}
\affiliation{Laboratory for Neutron Scattering and Imaging, Paul Scherrer Institut, 5232 Villigen,  Switzerland}

\author{T.~Fennell}
\affiliation{Laboratory for Neutron Scattering and Imaging, Paul Scherrer Institut, 5232 Villigen, Switzerland}

\author{D. Khalyavin}
\affiliation{ISIS Facility, Rutherford Appleton Laboratory, Chilton, Didcot, Oxon OX11 0QX, United Kingdom}

\author{Z. Yan}
\affiliation{Laboratory for Solid State Physics, ETH Z{\"u}rich, 8093 Z{\"u}rich, Switzerland}

\author{S. Gvasaliya}
\affiliation{Laboratory for Solid State Physics, ETH Z{\"u}rich, 8093 Z{\"u}rich, Switzerland}

\author{A. Zheludev}
\email{zhelud@ethz.ch}
\homepage{http://www.neutron.ethz.ch/}
\affiliation{Laboratory for Solid State Physics, ETH Z{\"u}rich, 8093 Z{\"u}rich, Switzerland}

\date{\today}

\begin{abstract}
The highly-frustrated rare-earth based magnet \NBWO is a promising candidate in the search for proximate spin liquid physics.  We present a thorough investigation on single crystals of this material using bulk and microscopic techniques.  Magnetization data reveal a fractional magnetization plateau for three different investigated field directions.  The magnetic phase diagram is mapped out from calorimetric data and exhibits several domes of magnetic order below 0.3 K.  Propagation vectors for all ordered phases are presented. The results suggest complex ordering in this material, and unveil the existence of a commensuration transition of the propagation vector at zero magnetic field.  A scenario where interplane exchange interactions are essential to a magnetic model of \NBWO is discussed.
\end{abstract}

\maketitle

\section{Introduction}

Strongly frustrated quantum antiferromagnets (AFM) are known to realize a panoply of magnetic states due to the delicate equilibrium between the magnetic interactions. In the presence of magnetic fields, the large ground state degeneracy is lifted in subtle and diverse ways, which leads to extremely rich phase diagrams. Realization of spin-density waves \cite{PhysRevLett.129.087201}, magnetization plateaus \cite{smirnov2007triangular,levy2008field} commensurate-incommensurate transitions \cite{wada1982incommensurate},  and even more exotic order like spin nematicity \cite{mourigal2012evidence,bhartiya2019presaturation} is not rare, particularly in quasi-low-dimensional systems.  

The archetypal model in 2D frustrated magnetism is the kagome lattice Heisenberg S = 1/2 AFM (KHAF).  The impossibility of satisfying all magnetic interactions in this lattice results in a macroscopic degeneracy of the ground state already at a classical level \cite{kermarrec2021classical}. Turning to S = 1/2 spins promotes the quantum fluctuations on the ground state giving rise to highly non-trivial phases \cite{anderson1973resonating}. Arguably, the most intriguing state is the hypothesized Quantum Spin Liquid (QSL) \cite{Zhou2017Quantum} ground state.  The prediction of fractionalization of quasiparticles in a 2D system triggered extensive effort from both theory and experimental perspectives \cite{savary2016quantum,chamorro2020chemistry}.  Nevertheless, the QSL phase remains elusive \cite{broholm2020quantum} as it constitutes a very fragile state. One of the main causes of the instability of the QSL states is the presence of  terms in the Hamiltonian that lift the ground state degeneracy \cite{bernu2013exchange, yoshida2022frustrated}.  The many different ways to lift this degeneracy have led to a flurry of new magnetic structures \cite{leblanc2013monte,yoshida2017unusual,okuma2019series,jeschke2019kagome}. However, occasionally deviations from a putative KHAF tend to stabilize QSL phases.  In particular, the so called \textit{breathing} anisotropy has been predicted to favor a resonance valence bond solid ground state for a wide range of coupling parameters \cite{iqbal2020gapped,jahromi2020spin}.  

In this context, the recently discovered family $R_3BWO_9$ of rare-earth antiferromagnets is an optimal platform for the search of spin-liquid candidates \cite{ashtar2020new}. Here $R$ is a trivalent rare-earth element and the large difference in size of the constituent atoms prevents anti-site chemical disorder.  All of the members of the family realize a breathing kagome lattice in their basal plane and show no sign of magnetic ordering down to 2 K. The strong spin-orbit coupling in combination with crystal electric field effects opens the possibility of realizing effective $J_{eff} =$ 1/2 magnetic moments. 

Among all compounds in the family,  the most promising is \NBWO. A large Weiss temperature \cite{ashtar2020new} has been reported and the total angular momentum of Nd$^{3+}$ ($J$ = 9/2) makes it a Kramers-doublet system.  No magnetic long-range order has been found in previous studies down to 1.8 K.  However,  little is known so far about its magnetism.  In this study we report on the low temperature properties of single crystals of \NBWO. We found static magnetic long-range order below 0.3 K.  The observed magnetism suggests a three dimensional network of exchange interactions. Nonetheless, due to the highly frustrated interaction a complex phase diagram is realized. 

The paper is structured as follows. First, a summary of the various methods used is provided. Then, we outline the main results of the experiments.  Subsequently, a detailed discussion of the main outcome is provided, including a thorough description of the magnetic structure and a detailed picture of the magnetic phase diagram under applied fields. Finally, the main conclusions are drawn and further steps in the search of QSL physics are examined. 

\section{Methods}

\begin{figure}[tbp]
\includegraphics[scale=1]{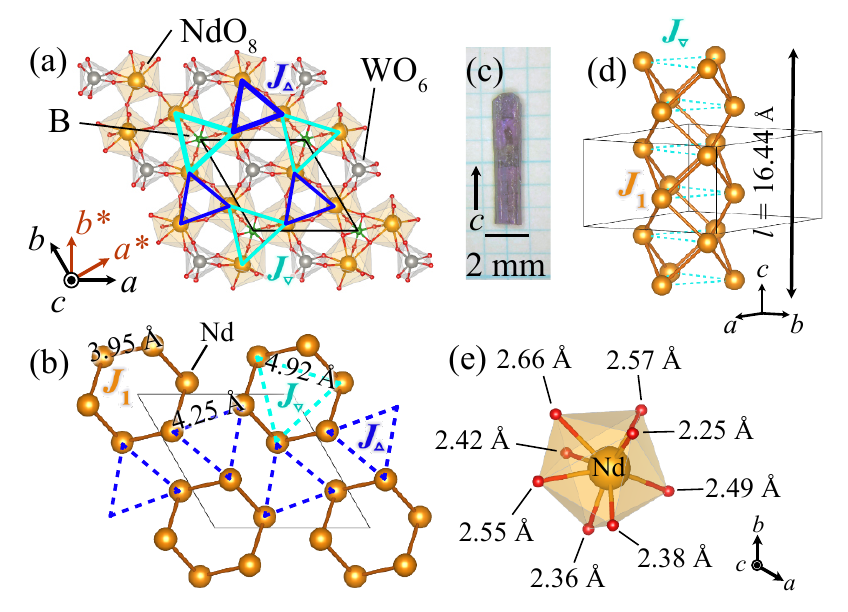}
\caption{Crystal structure and superexchange topology in \NBWO. (a) Schematic structure reflecting the purported kagome interaction in the crystallographic \textit{ab} plane.  Only atoms with 0 $\le z\le$ 0.5 are shown here. There is an additional kagome plane displaced by half lattice parameter along the \textit{c} crystallographic direction.  (b) The shortest superexchange Nd-O-Nd bond links neodymium atoms in different kagome planes, forming isolated \textit{spin tubes} along the \textit{c} axis arranged in a triangular lattice. The kagome bonds are shown for reference along with bond distances.  (c) A typical single crystal sample of \NBWO. (d) A single spin tube is  unfrustrated. However, further-neighbor interactions frustrate the system. An arrow indicates the size of the magnetic supercell at zero field.  (e) The environment of neodymium has very low symmetry,  resulting in a $C_1$ point group for the magnetic ion.  Nd-O distances are indicated.}
\label{fig:Crystal_structure}
\end{figure}

\NBWO crystallizes in a hexagonal structure, with space group $P6_3$ (No. 173), where the magnetism stems from the effective magnetic moment of the Nd$^{3+}$ ions. Single crystal samples were grown by spontaneous crystallization using a flux method as described in \cite{majchrowski2003growth}. Purple transparent single crystals with well defined facets were obtained [Fig.~\ref{fig:Crystal_structure}(c)]. Typical masses range from a few micrograms to 40 mg and different samples were used in this study, depending on the technique.  The chemical structure of the different single-crystal samples used in this study was validated using single-crystal X-ray diffraction on a Bruker APEX-II instrument, and was found to be in agreement with previous reports \cite{ashtar2020new}.  The structure is schematically depicted in Fig.~\ref{fig:Crystal_structure}, where the kagome-lattice bonds can be readily identified.  Powder samples of \NBWO , as well as of the non-magnetic La$_{3}$BWO$_9$, were synthesized by a solid state reaction. The correct chemical structure and the quality of the powders was checked with powder X-ray diffraction in a Rigaku MiniFlex diffractometer. Boron-11 enriched samples (both powder and single crystals) were also prepared for their use in neutron scattering experiments.

Measurements of heat capacity,  magnetocaloric effect (MCE),  magnetization and magnetic torque were carried out using a $^3$He-$^4$He dilution refrigerator insert for the Quantum Design Physical Property Measurement System (PPMS).  A sample of mass 0.131 mg was used for both heat capacity and MCE measurements. Heat capacity data were collected using a standard relaxation method from Quantum Design for temperatures $100$ mK $<$ \textit{T} $<$ 4 K in applied fields of 0 T $<$ $\mu_0$\textit{H} $<$ 3 T.  The magnetic field was applied along the crystallographic \textit{a$^*$}, and \textit{c} directions.  In zero field, data were collected from 100 mK to 300 K.  Heat capacity data of La$_{3}$BWO$_9$  were measured down to 2 K and extrapolated to lower temperatures from an empirical fit to a $T^3$-power law.  MCE data were measured using the same puck as for heat capacity. The change of temperature of the sample was recorded as the magnetic field was swept up and down at a constant rate. In order to avoid self heating of the puck, the field change rate was optimized and a value of 0.5~mT/s was selected. In the terminology of MCE measurements, our experiment was conducted under \textit{equilibrium} conditions. 

Magnetization was measured using an in house made Faraday-balance capacitive magnetometer \cite{blosser2020miniature} at 120 mK and 2 K and magnetic fields applied along three orientations: \textit{a$^*$},  and \textit{b}, and \textit{c}.  Additional measurements of magnetization carried out in the MPMS system at 2 K were used to calibrate the low temperature data and obtain absolute units (not shown here).  Using the same setup, magnetic torque was measured up to 3 T and for temperature from 120 mK to 600 mK.  The torque data correspond to the deflection of a small cantilever on which the sample is mounted. The magnetic field sweeping rate was also optimized to minimize heating due to eddy currents. 

Magnetic susceptibility was measured using the Quantum Design Magnetic Property Measurement System (MPMS) SQUID Magnetometer. The temperature range from 1.8 K to 300 K was probed using a small polarizing field applied along three crystal directions: \textit{a$^*$},  and \textit{b}, and \textit{c}.  The probing field was $\mu_0H = 0.1$ T, where $\mu_0$ denotes the permeability of vacuum. 

Inelastic neutron scattering on powder samples of \NBWO was measured to investigate the crystal electric field induced scheme of total angular momentum states. The instrument of choice was the thermal neutron triple-axis-spectrometer EIGER at PSI.  11.1 g of \NBeWO was sealed in an aluminum can and installed in a standard $^4$He orange cryostat.  A final wavelength of $k_f$= 2.66 \AA$^{-1}$ ($\lambda$ = 2.36 \AA) was chosen, using a pyrolytic graphite filter to eliminate higher-order neutrons without further collimation.  Data were measured at constant scattering angle,  $2\theta$. The background was investigated to select the optimal value for the scattering angle,  sufficiently far from the direct beam and low enough to have good counting and small decay in the signals due to magnetic structure factors.  A value of $2\theta=10^\circ$ was chosen, and the incident energy was scanned at three different temperatures: 1.5 K, 100 K and 300 K. 

Neutron single crystal diffraction was used to investigate the magnetic structures in the ordered phases. A single crystal sample of 18 mg in mass of  \NBeWO and 5.5$\times$1.4$\times$0.8 mm$^3$ was studied using two different instruments. Measurements with $H\parallel \textit{a}^*$ were carried out at the Thermal Single Crystal Diffractometer ZEBRA at the Swiss Spallation Neutron Source, SINQ,  in the Paul Scherrer Institut (PSI, Switzerland). The diffractometer was used in conjunction with a $^3$He-$^4$He dilution refrigerator and a 6-T magnet. The crystal was aligned with its $\textit{a}^*$ axis vertical,  the same direction as the applied magnetic field.  Neutron wavelengths of $\lambda$ = 2.314 \AA\xspace and 1.383 \AA\xspace were selected,  provided by the PG(200) and Ge(220) monochromators.  Additional measurements with $H\parallel \textit{c}$ were carried out in the time-of-flight diffractometer WISH at the ISIS facility in the Rutherford Appleton Laboratory, in the United Kingdom. The sample was mounted with its $c$ axis vertical and parallel to the magnetic field. A $^3$He-$^4$He dilution refrigerator and a 10-T magnet were used to access the ordered states in \NBWO. 

\section{Experimental results}

\subsection{Magnetic susceptibility}

\begin{figure}[tbp]
\includegraphics[scale=1]{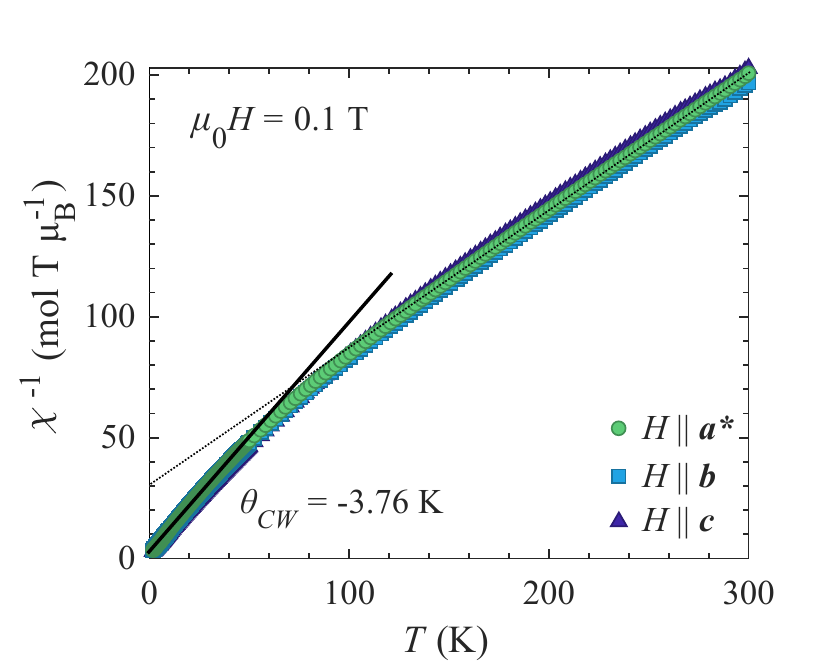}
\caption{Inverse magnetic susceptibility on single crystals. Data show measurements for three field orientations.  A small probing field of 0.1~T was used for all measurements.  The black solid line represents a Curie-Weiss model with the average Weiss temperature and effective moment parameters, given in Table.~\ref{table:Curie}. }
\label{fig:Susceptibility}
\end{figure}

Figure~\ref{fig:Susceptibility} shows inverse susceptibility measurements for probing fields applied along the crystallographic directions \textit{a}$^*$\textit{b}, and \textit{c}.  Down to the lowest accessible temperature of 1.8 K, these data show no sign of magnetic ordering. 

A fit of the experimental data to a Curie-Weiss model is shown overlaid on the experimental results.  A good agreement is found for data above 130 K, with a large, negative Weiss temperature. The resulting Weiss temperatures, $\theta_W$ are given in Table.~\ref{table:Curie}, as well as the corresponding effective magnetic moments extracted from the Curie constants as $C=N_A\mu_0\mu_{eff}^2/(3k_B)$.  The obtained effective magnetic moments are close to the value expected for a free Nd$^{3+}$ ion: $\mu_{eff} = g_J \sqrt{J(J+1)} \mu_B = 3.6 \mu_B$.  Importantly, the susceptibility data show little dependence on the direction of the magnetic field, which suggests that, the resulting magnetic anisotropy remains quite small.  Our results are consistent with those reported in Ref.\cite{ashtar2020new} on polycrystal samples. 

\begin{table}
		\caption{Fitting parameters from the Curie-Weiss model for data shown in Fig.	~\ref{fig:Susceptibility}.}
  		\centering 
  \begin{threeparttable}
   		 \begin{tabular}{c|c c | c c}
   		  \cmidrule(l  r ){2-5}
     	& \multicolumn{2}{|c|}{\text{ 200 K $\leq$ \textit{T} $\leq$ 300 K}} & \multicolumn{2}{|c}{\text{20 K $\leq$ \textit{T} $\leq$ 60 K}} \\
      \cmidrule(l  r ){2-5}
 	&  $\theta_W$ (K) & $ \mu_{eff} $ ($\mu_B$) & $\theta_W$ (K) & $ \mu_{eff} $ ($\mu_B$)  \\
   \midrule  \midrule
    		$\textbf{H}\parallel \textit{a}^*$ & -54.3 &  3.76 &  -3.78 &  2.94\\ 
    		\cmidrule(l  r ){1-5}
		$\textbf{H}\parallel \textit{b}$& -54.7 &  3.79 &  -3.82 &  2.90\\ 
		\cmidrule(l  r ){1-5}
		$\textbf{H}\parallel \textit{c}$ & -59.2 & 3.77 &   -3.68 & 2.91\\ 

    \midrule\midrule
    \end{tabular}
    
    \end{threeparttable}
    \label{table:Curie}
\end{table}

Below 130 K a clear deviation from the high temperature fit is observed. This is roughly consistent with the existence of a crystal electric field (CEF) level at 15.9 meV (see below), signaling the total depletion of the population of the first excited state.  A Curie-Weiss analysis is heavily affected by the partial population of excited multiplets and lead to an overestimation of exchange parameters and exchange couplings.  Therefore, an additional fit to a Curie-Weiss law for a temperature range far enough from the CEF resonance has been performed. The  results are also summarized in Table ~\ref{table:Curie}. Temperatures in the range between 20 K and 60 K were considered for this fit.  The resulting Weiss temperatures are much reduced compared to the high temperature fit. However, they still reflect a predominant antiferromagnetic interaction in \NBWO. The effective magnetic moments are also reduced with respect to their high temperature value, yielding an average moment of $\mu_{eff} =2.92 \mu_B$. 

\subsection{CEF level scheme}

\begin{figure}[tbp]
\includegraphics[scale=1]{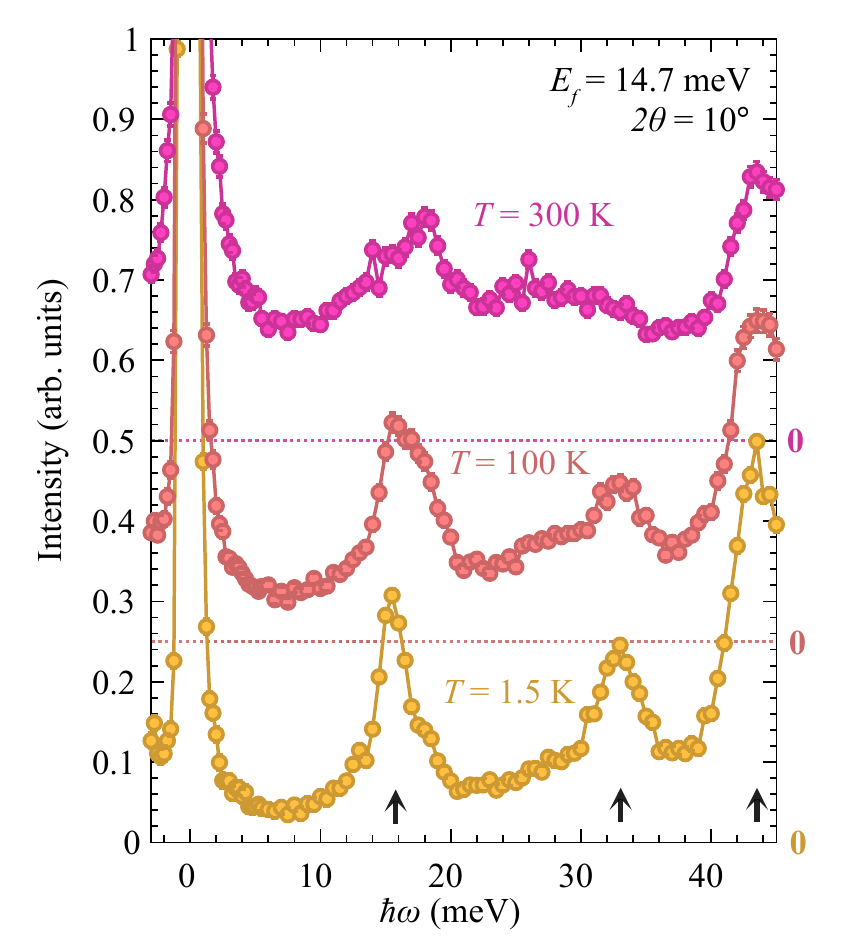}
\caption{Inelastic neutron scattering intensity at a constant scattering angle for three different temperatures.  The final energy of $E_f$= 14.7 meV was fixed and incident energy varied, fixing a 10 degree scattering angle. CEF resonances are indicated by black arrows. An offset of 0.25 and 0.50 units was added for visibility,  a dashed line indicates the reference zero for those data sets. }
\label{fig:Neutron_CEF}
\end{figure}

The inelastic neutron scattering spectra are shown in Fig.~\ref{fig:Neutron_CEF}.  Large intensity at zero energy transfer corresponds to quasielastic scattering. Three  resonances are identified at 15.9, 32.8, and 43.7 meV, which we ascribe to CEF induced levels due to their temperature dependence.   Importantly, no resonance is found below 15.9 meV. Since the total angular momentum $J = 9/2$ of the free Nd$^{3+}$ is expected to be fully split into five Kramers doublets, this suggests that the low temperature physics of \NBWO can indeed be described in terms of the lowest laying doublet, giving rise to an effective two-level system well below $\Delta =15.9$ meV $\approx 180$ K. 

\subsection{Specific heat}

\begin{figure}[tbp]
\includegraphics[scale=1]{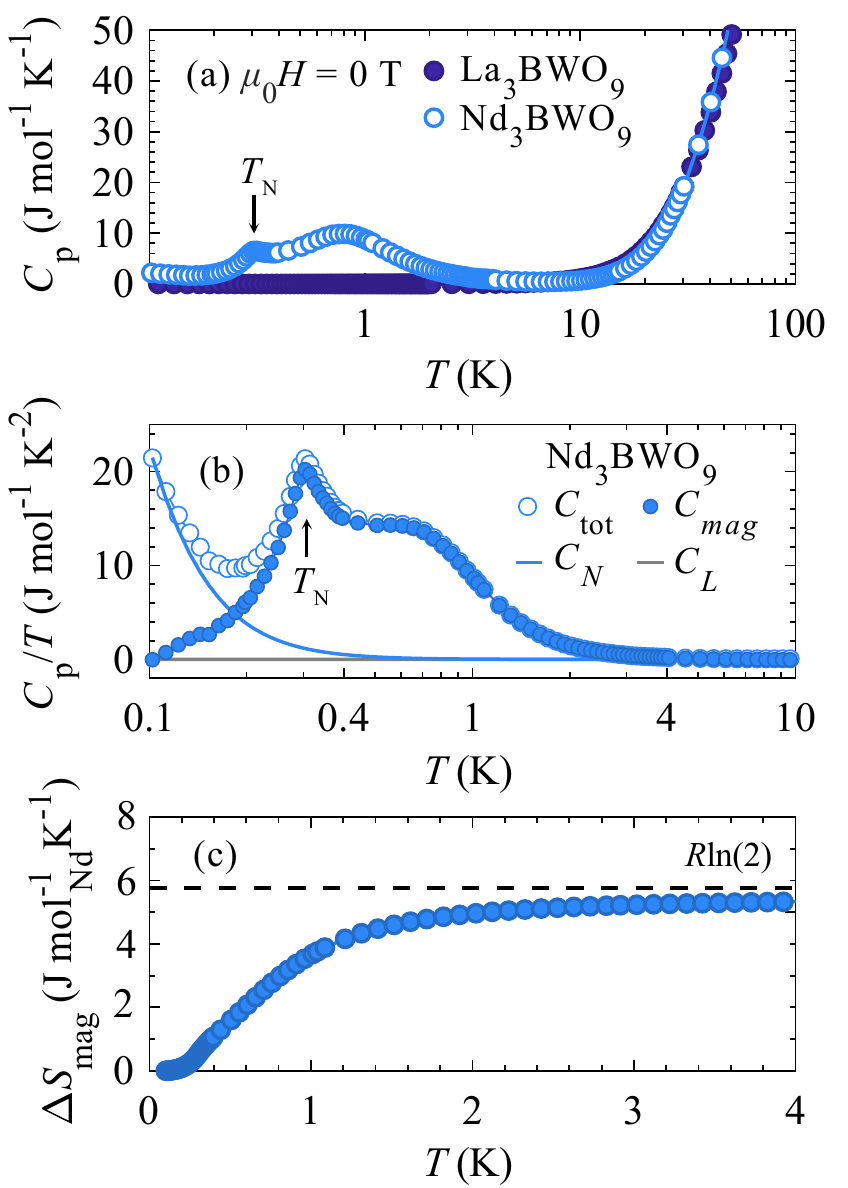}
\caption{(a) Total specific heat at zero magnetic field for Nd$_{3}$BWO$_9$ and the nonmagnetic isostructural compound La$_{3}$BWO$_9$.  Nd$_{3}$BWO$_9$ shows a substantial magnetic contribution to specific heat below 3 K. (b) Total specific heat (open circles) and magnetic specific heat (filled circles) after subtraction of lattice and nuclear degrees of freedom. Lattice ($C_L$) and nuclear ($C_N$) contribution are estimated as discussed in the text. A lambda anomaly can be found at $T_N =$ 0.30 K, signaling the onset of long-range magnetic order. (c) The magnetic entropy per Nd$^{3+}$ ion saturates above 3 K. A dashed line represents the expected value for a two-level system at infinite temperature.  }
\label{fig:HC_Zero_field}
\end{figure}

\begin{figure}[tbp]
\includegraphics[scale=1]{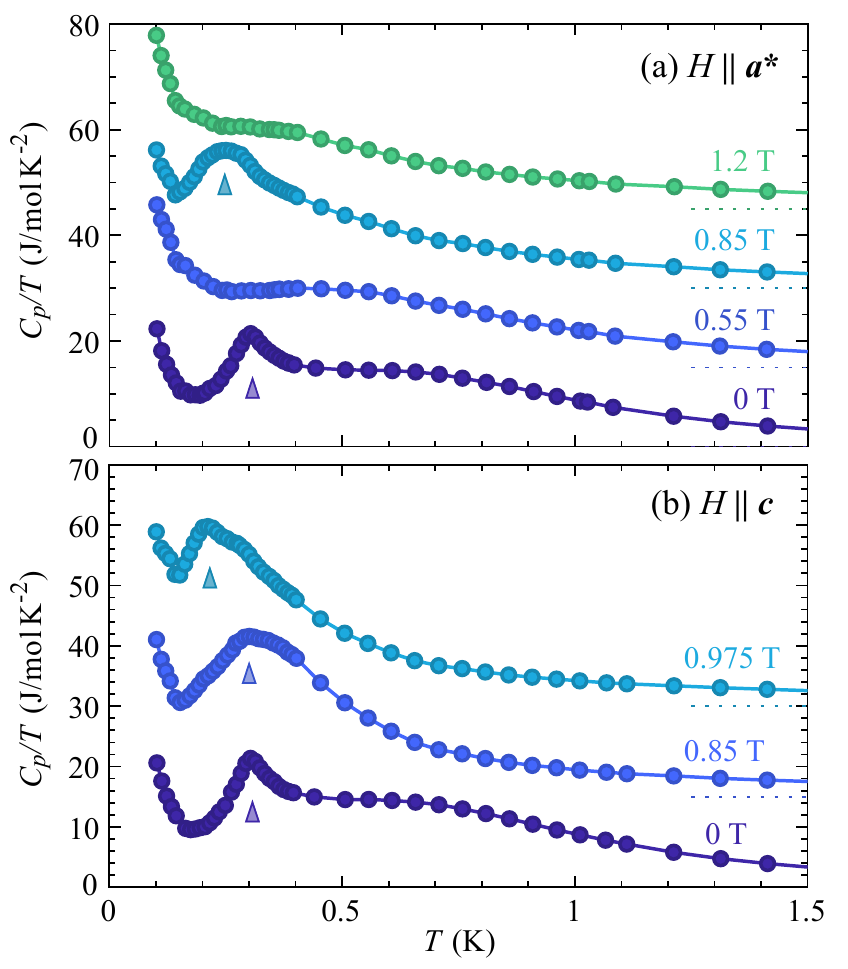}
\caption{Typical temperature scans of specific heat for different fixed values of magnetic field applied along (a) $\textbf{H}\parallel\textit{c}$ and (b) $\textbf{H}\parallel\textit{a}^*$. An offset of 15 J/mol/K$^2$ has been added for visibility.  Solid filled triangles show features associated with the phase transitions discussed in the main text.}
\label{fig:HC_Tscan}
\end{figure}

\begin{figure}[tbp]
\includegraphics[scale=1]{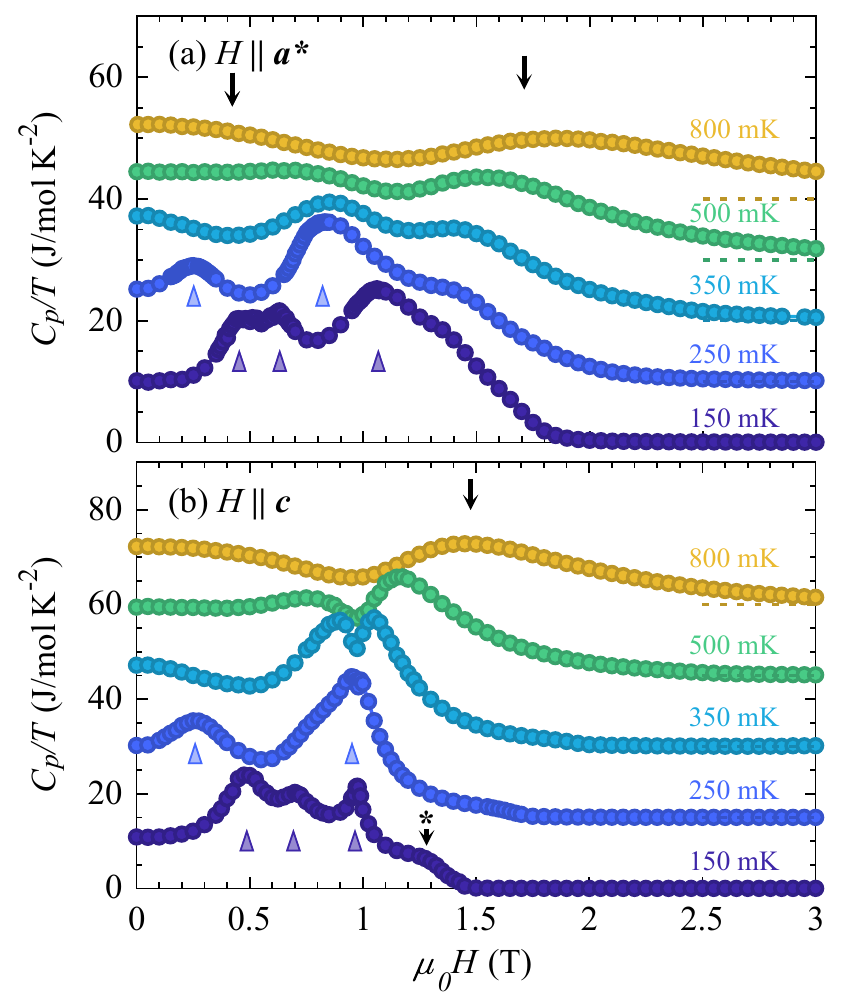}
\caption{Typical field scans of specific heat measured at constant temperature in \NBWO for (a) $\textbf{H}\parallel\textit{c}$ and (b) $\textbf{H}\parallel\textit{a}^*$. An offset of 10 or 15 J/mol/K$^2$ is added for visibility between the scans for (a) and (b), respectively. Solid filled triangles show features associated with the phase transitions discussed in the main text. Black arrows signal the existence of broad double-hump features,  described in the text.  An asterisk shows a feature above the saturation transition.}
\label{fig:HC_Hscan}
\end{figure}

Specific heat as a function of temperature and magnetic field is used to unveil the magnetic phase diagram of \NBWO at ultra-low temperatures.  Data obtained at zero field are shown in Fig.~\ref{fig:HC_Zero_field}.  \NBWO shows an upturn in specific heat below 4 K with two clearly distinct features [Fig.~\ref{fig:HC_Zero_field}(a)]. Around 1 K, a hump in specific heat suggests the onset of short-range magnetic correlations \citep{dutton2012quantum}. At $T_N =$ 300 mK we found a sharp lambda anomaly representing the transition into magnetic long range order.  Below $T_N$ the specific heat signal remains large down to the lowest accessible temperatures in our setup, likely due to nuclear specific heat from the rare-earth ions. In order to understand exactly the nature of the magnetic specific heat, we have examined the different contributions and subtracted them from the measured total specific heat.  

To estimate the phononic contribution, we synthesized the non-magnetic isostructural material La$_{3}$BWO$_9$ and measured its specific heat in the same range of temperatures. This is shown in Fig.~\ref{fig:HC_Zero_field}(a) and represents the lattice contribution, $C_L$, in Fig.~\ref{fig:HC_Zero_field}(b). 

An accurate estimation of the nuclear contribution to specific heat is usually much more complicated, as a variety of  effects has to be considered. These include dipole and quadrupolar splitting, or hyperfine coupling between nuclei and electrons (which can be quite significant in magnetically ordered materials). Neodymium has two isotopes with nonzero dipolar and quadrupolar momenta, out of its 7 stable isotopes.  Following the reasoning in Ref.~\cite{kimura2013quantum}, the effect of quadrupolar splitting is assumed to be small compared to that of hyperfine coupling, and we neglect it here. In a magnetized phase,  local fields are expected to be sizable and therefore hyperfine coupling may significantly contribute to specific heat.  The contribution from dipole field splitting from a single isotopic species is given by

\begin{multline}
	C_{H,i} =\\
 	N_Ak_B \frac{\alpha^2_i}{4I_i^2}   \left[\frac{1}{\sinh^2\left(\frac{\alpha_i }{2I_i}\right)}-\frac{(2I_i+1)^2}{\sinh^2\left(\frac{(2I_i+1)\alpha_i }{2I_i}\right) }\right]
\end{multline}
\label{Eq:Nuclear_HC}
where $\alpha_i = A_{H}(\mu_{Hyp}^{Nd}/g_J)I_i/k_BT$, and $I_i$ is the nuclear spin, $g_J = 8/11$ (Landé factor for Nd), $N_A$ is the Avogadro constant, and $k_B$ the Boltzmann constant. $A_{Hyp}$ represents the strength of the hyperfine coupling and here we made a second approximation. We assume all the nuclei couple equally to the electron density and the value of $A_{Hyp}$ is approximated as that of Nd metal  \cite{bleaney1963hyperfine}.  $\mu_{Hyp}^{Nd}$ denotes the static dipole moment of the Nd$^{3+}$ ions. This is precisely the origin of the local field and for its value we chose the averaged effective magnetic moment from the magnetic susceptibility data at low temperatures $\mu_{Hyp}^{Nd} = 2.914\mu_B $. Finally, the different species are summed, weighted by their isotopical abundance to obtain the temperature dependence of nuclear specific heat. 

This model with no free parameters is in excellent agreement with the lowest temperature data, as shown in Fig.~\ref{fig:HC_Zero_field}(b).  Having modeled the nuclear specific heat, the magnetic specific heat can be extracted by subtraction.  The magnetic specific heat was subsequently integrated to obtain the temperature dependence of magnetic entropy, depicted in Fig.~\ref{fig:HC_Zero_field}(c). The high temperature trend of this quantity approaches the value of $R\ln(2)$,  the expected value of a two-level system. 

In a magnetic field, a simple estimation of the contribution of the nuclear spin due to Zeeman splitting could not account for the effects observed here.  Low temperature data in Fig.~\ref{fig:HC_Tscan} show that the effect of nuclear specific heat is of the same order of magnitude up to 1.2 T and it is not strongly field dependent.  This suggests that also in a field the main contribution comes from hyperfine coupling. However, a quantitative determination of this effect under magnetic fields becomes paramount. 

The evolution of the specific heat of \NBWO under magnetic fields is shown in Fig.~\ref{fig:HC_Hscan} for fields along two different crystallographic directions. The total heat capacity is displayed here, without subtraction of lattice or nuclear degrees of freedom.  Typical-field scans show a number of anomalies that are consistent with the existence of three different phases with static magnetic order at low temperatures. 

Up to three distinct features can be observed for $\textbf{H}\parallel\textit{a}^*$ at the lowest temperature, at 0.45, 0.62 and 1.05 T and are marked with triangles in Fig.~\ref{fig:HC_Hscan}(a).These features are rather spread in fields, specially at saturation.  However, the existence of thermodynamic transitions has been confirmed by neutron diffraction (as discussed below). The two lower field anomalies move apart as the temperature is increased.  The two higher field anomalies merge at 0.25 K, denoting the highest temperature of the ordered phase. Though the specific heat anomalies in Fig.~\ref{fig:HC_Hscan}(a) are too broad for a precise estimation of the upper critical field,  this quantity can be deduced from magnetocaloric effect measurements (see below).

For fields orthogonal to the hexagonal plane ($\textbf{H}\parallel\textit{c}$) at the lowest temperature one finds two anomalies at 0.5 T, 0.8 T and a sharper one at 0.95 T. [Fig.~\ref{fig:HC_Hscan}(b)]  Notably, in this configuration the different anomalies appear narrower than for $\textbf{H}\parallel\textit{a}^*$, especially at the saturation field. The first two anomalies move apart as the temperature is increased, while the higher field anomaly barely shifts in position up to 0.2 K.  The low field anomaly shifts towards zero field and disappears as $T_N$ is reached.  The two high-field anomalies merge at $T$ = 0.2 K.  From the high field anomaly we extract an estimate of the saturation field of $\mu_0H_c = 0.975(3) $ T.  Interestingly, an extra feature can be identified above saturation (asterisk in Fig.~\ref{fig:HC_Hscan}(a)).  This feature shifts to higher fields as the temperature is increased and decreases rapidly in magnitude. Above 0.2 K it is hardly identifiable. 

Finally, double-hump features can be observed above 0.3 K for both magnetic field configurations.  These are significant up to the highest measured temperatures and particularly prominent around the saturation field (black arrows in Fig.~\ref{fig:HC_Hscan}).  For $\textbf{H}\parallel\textit{c}$ the amplitude of these modulations is larger than in $\textbf{H}\parallel\textit{a}^*$.  Such features are often associated with a low-dimensional crossover from the zero field disordered phase to the fully polarized state without the occurrence of a phase transition \cite{Ruegg2008Thermo,Blosser2018Quantum,Hayashida2019One,Flavian2020Magnetic}.  

\subsection{Magnetocaloric effect}

\begin{figure}[tbp]
\includegraphics[scale=1]{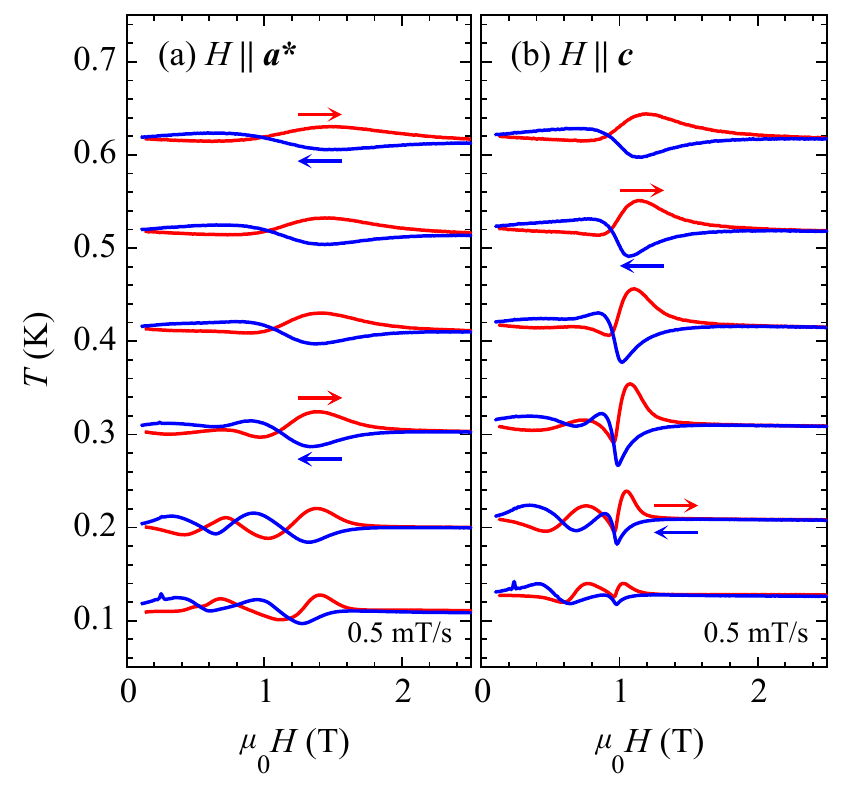}
\caption{Plots of the magnetocaloric effect in \NBWO for different base temperatures and fields applied along (a) $\textbf{H}\parallel\textit{a}^*$ and (b) $\textbf{H}\parallel\textit{c}$. For all the scans, red (blue) color represents data measured while driving the magnetic field up (down).  A ramping rate of 0.5 mT/s was used throughout all the measurements. Small prominent features (specially at low fields) are spurious and the result of an unstable platform. }
\label{fig:MCE}
\end{figure}

\begin{figure}[tbp]
\includegraphics[scale=1]{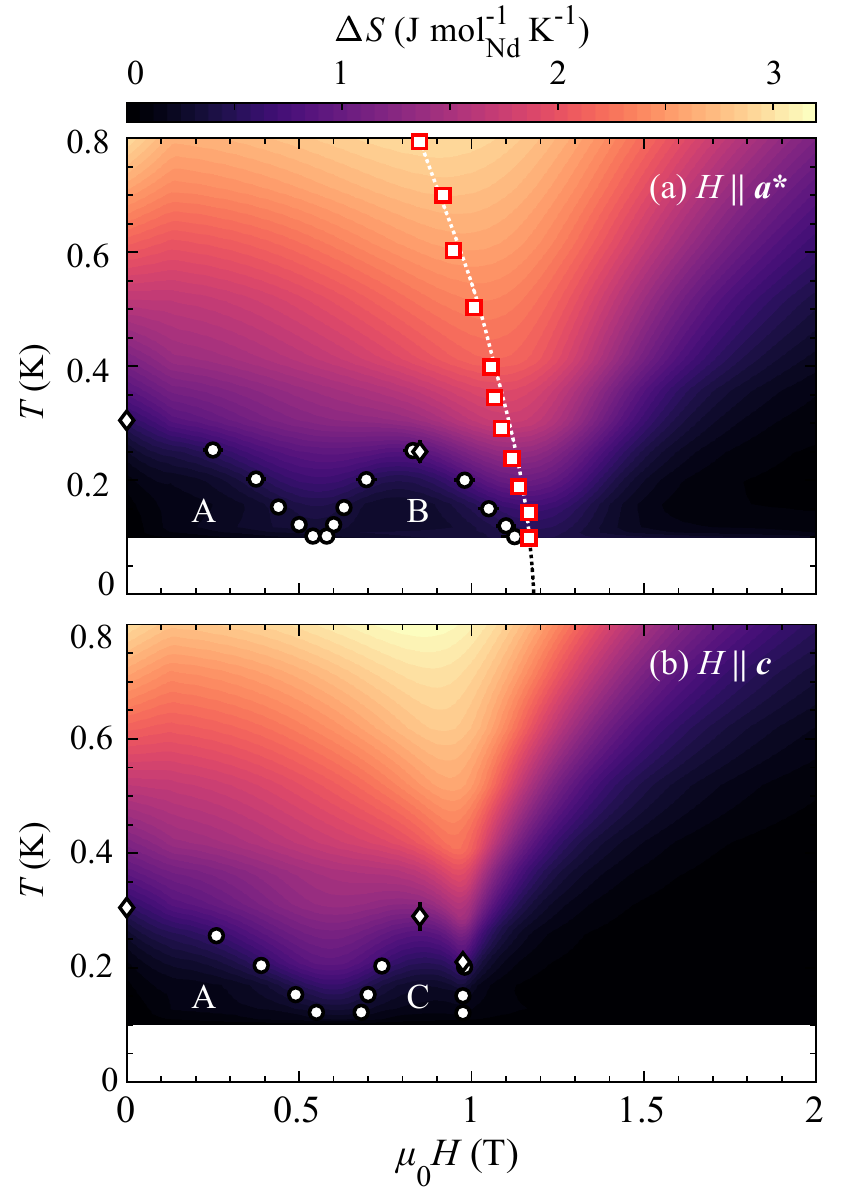}
\caption{Entropy maps in false color for two magnetic field orientations. In false color plots, the change in entropy extracted from magnetocaloric data from Fig.~\ref{fig:MCE}. Filled circles (diamonds) denote phase anomalies associated with phase transitions from specific heat field (temperature) scans for (a) $\textbf{H}\parallel\textit{a}^*$ and (b) $\textbf{H}\parallel\textit{c}$.  In (a) red squares show the maxima of entropy at the measured temperatures.  A white dashed line is a power law fit to the data showing the best estimate for the upper critical  field.}
\label{fig:MCE_entropy}
\end{figure}

Magnetocaloric effect (MCE) measurements in \NBWO provide key information on the nature of the various phase transitions found with other techniques  \cite{zapf2014bose,kohama2010ac,kohama2022high}.  Representative temperature profiles are summarized in Fig. ~\ref{fig:MCE}.  Several crossings can be observed for both configurations.  The observed anomalies are too broad to assign exactly a transition point. Due to the proximity of the thermodynamic transitions in the phase diagram, features corresponding to both transitions merge and overlap.  In our measurements the field is swept slow enough as to ensure equilibrium conditions. 

Data measured with $\textbf{H}\parallel\textit{a}^*$ show mostly symmetric features around the crossing points. Particularly, this suggests that the measured phase transitions are of second order.  In contrast,  the low temperature profiles for $\textbf{H}\parallel\textit{c}$ show two distinct behaviors. At 0.6 T one finds a roughly symmetric feature, suggesting again a second order phase transition. This is different at 0.975 T, where a very asymmetric feature appears, pointing to a first order or discontinuous transition.  

Finally, the absence of anomalies above the saturation field for $\textbf{H}\parallel\textit{c}$ must be noted. The features observed in the fully polarized phase in Fig.~\ref{fig:HC_Hscan}(b) leave no trace in the MCE data in the same configuration. 

The MCE technique is based on the change of entropy in a magnetic system as it is driven through a phase transition, crossover,  level crossing, etc. Consequently, one can retrieve the change in entropy in a system from the change in temperature against magnetic field \cite{zapf2014bose}.  Under equilibrium conditions, we obtain the entropy as 
\be
\Delta S = S(H) - S_0 = - \int \kappa \frac{T-T_{bath}}{T}dt 
\ee

where $\kappa$ is the thermal conductivity of the thermal link in the calorimeter, $T$ is the sample temperature, and $T_{bath}$ is the thermal bath temperature.  Integration of the data in Fig.~\ref{fig:MCE} gives rise to the entropy maps displayed in Fig.~\ref{fig:MCE_entropy}.  The data above 0.2 K are a good picture of the entropy stored in the magnetic subsystem.  However, for temperatures below 0.15 K imperfect equilibrium conditions prevent a reliable estimation of entropy.  A strong accumulation of entropy is observed above the saturation transitions for both field configurations. The position of the peaks in entropy match the estimated position of the critical fields from specific heat. For $\textbf{H}\parallel\textit{a}^*$, the maxima in entropy at different temperatures were used to obtain an accurate estimate of the upper critical field. A fit to the data provides $H_{c,a^*} = 1.187(13)$ T. This value is consistent with the various probes used in this study. 

\subsection{Magnetization}

\begin{figure}[tbp]
\includegraphics[scale=1]{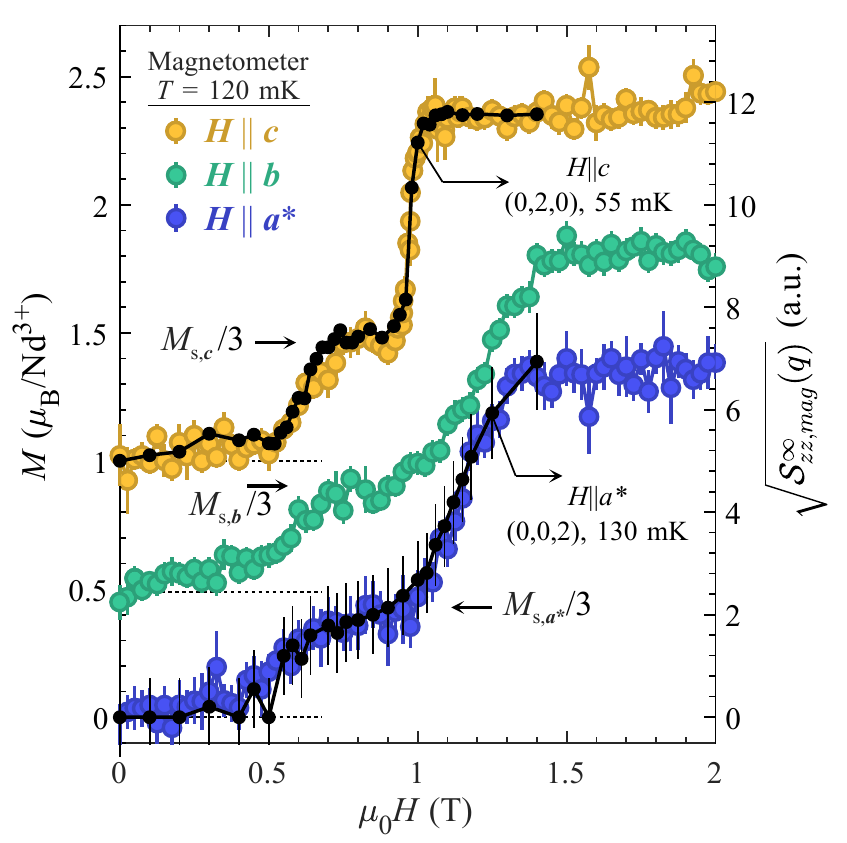}
\caption{Magnetization per Nd$^{3+}$ ion measured at 120 mK in \NBWO for magnetic fields along the crystallographic directions \textit{a$^*$}, \textit{b}, and  \textit{c} from bulk measurements (left axis). Magnetization extracted from neutron diffraction intensity of nuclear reflections is superimposed to the corresponding bulk data.  Plotted is the rescaled square root of the static, magnetic structure factor $\mathcal{S}_{z,z}^\infty(\textbf{q})$ (right axis). The measured reflections ($\textbf{Q}$) are indicated in the figure. Two of the data sets have been offset vertically by 0.5 and 1.0 units to improve visibility (a dashed line indicates their respective zero). The magnetization value at 1/3 of saturation is indicated for each individual data set by an arrow next to the plateau state.  }
\label{fig:Magnetization}
\end{figure}

The evolution of magnetization under a magnetic field provides insight on the type of order in \NBWO. Strikingly, a fractional magnetization plateau is observed for all measured configurations, as displayed in Fig.~\ref{fig:Magnetization}. The value of magnetization is consistent with a  fractional $m$=1/3 plateau and spans a range of fields of 0.2-0.3 T.  In addition, the zero field phase shows zero magnetization for all applied fields, which indicates the realization of a gapped phase at $T$ = 0.  Magnetization data for inequivalent directions in the hexagonal plane show very similar behavior, but differ from the results perpendicular to the plane.  

For $\textbf{H}\parallel\textit{a}^*$ and $\textbf{H}\parallel\textit{b}$ the zero magnetization phase extends up to 0.4 T.  Above 0.5 T the system transitions into the factional magnetization plateau state up to a marked limit at 1 T.  The transition into the fully saturated phase is gradual between 1 T and 1.3 T. 

In contrast, much sharper features are found when fields $\textbf{H}\parallel\textit{c}$ are applied.  A non-magnetizable phase appears up to 0.5 T, above which the system jumps rapidly into the plateau state at 0.65 T. The plateau terminates in a first-order jump to saturation around 1 T.  Notably, despite the presence of a first-order transition, our measurements did not show signatures of hysteresis across the saturation transition for $\textbf{H}\parallel\textit{c}$.

Saturation fields extracted from magnetization data are consistent with those found in the specific heat data. The values for saturation magnetization show little anisotropy,  finding 1.34(6) $\mu_B$, 1.31(3) $\mu_B$, and 1.35(4) $\mu_B$ per magnetic ion for configurations $a^*$, $b$, and $c$ respectively. It suggests a nearly isotropic g-tensor in the material. 

\subsection{Magnetic torque}

\begin{figure}[tbp]
\includegraphics[scale=1]{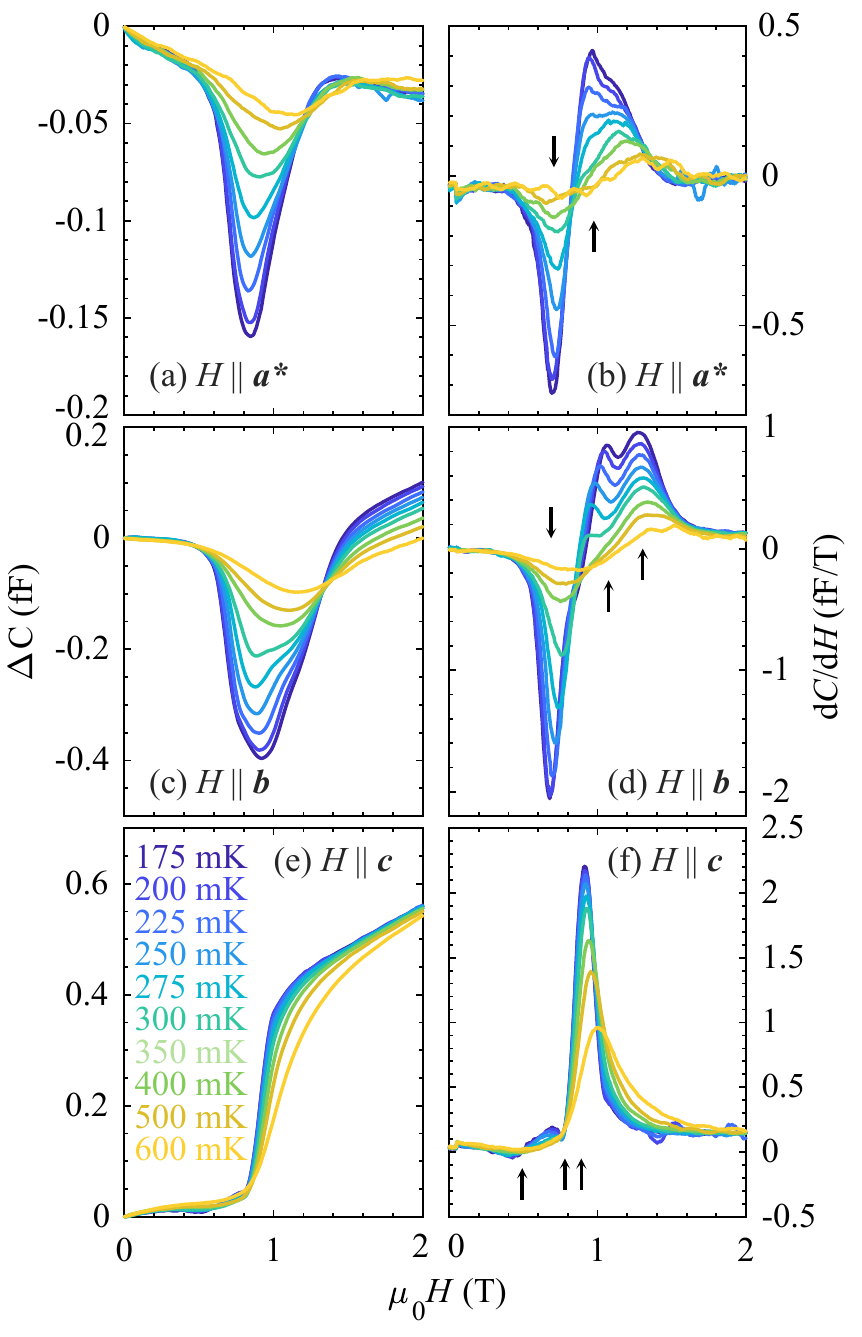}
\caption{Magnetic torque ($\Delta C$) and its field derivative measured at constant temperatures against magnetic field, for three field orientations: (a,b) \textit{a}$^*$, (c,d) \textit{b},  and (d,e) \textit{c}. For all data sets, a reference value of capacitance at zero field has been chosen and subtracted.  Black arrows indicate features that may be identified with phase transitions.}
\label{fig:Torque}
\end{figure}

Magnetic torque is arguably the most sensitive technique to magnetic phase transitions.  Raw data are presented as the change in the measured capacitance $\Delta C = C(H) - C(H=$ 0 $T)$  as a function of magnetic field for each temperature (Fig.~\ref{fig:Torque}).  The torque data show strong differences between the measurements in the basal plane and perpendicular to it, but the obtained results are very similar for both measurements within the plane. The raw data show some structure, but not sharp features as is customary in such measurements.  Phase transitions are best captured in the first derivative of the raw data ~\ref{fig:Torque}.  

Field derivative data show features that correspond with transitions observed in the other techniques reported in this study.  Direct comparison with the other data sets is necessary to pinpoint what anomalies represent real phase transitions. These features are indicated with arrows in the field derivative data [Fig.~\ref{fig:Torque}(b), \ref{fig:Torque}(d), and \ref{fig:Torque}(f)]. For $\textbf{H}\parallel\textit{a}^*$ two distinct anomalies can be observed in the scan at 175 mK,  at 0.68 T and at 0.99 T. These correspond to the lower and upper boundaries of the plateau phase. Data for $\textbf{H}\parallel\textit{b}$ show three anomalies at 0.68 T, 1.02 T and 1.28 T. The lower fields correspond again to the boundaries of the plateau phase.  Notably, these two anomalies come together as the temperature is increased and disappear above 300 mK. The higher field anomaly, which is broader and less sharp, corresponds to the crossover into the fully saturated state. Finally, fields applied along the \textit{c} direction reveal a completely different structure. Three anomalies can be identified at 0.48 T, 0.71 T, and 0.95 T.  The associated transitions in this case are the boundaries of the plateau for the high field features and the transition from the low field phase to paramagnet for the low field anomaly.  The low field features, though weak,  fade away as the transition temperature is overcome.  The high field anomaly remains up to the highest temperatures representing the crossover of the system into the fully polarized pseudospin. 

\subsection{Neutron diffraction}

\begin{figure}[tbp]
\includegraphics[scale=1]{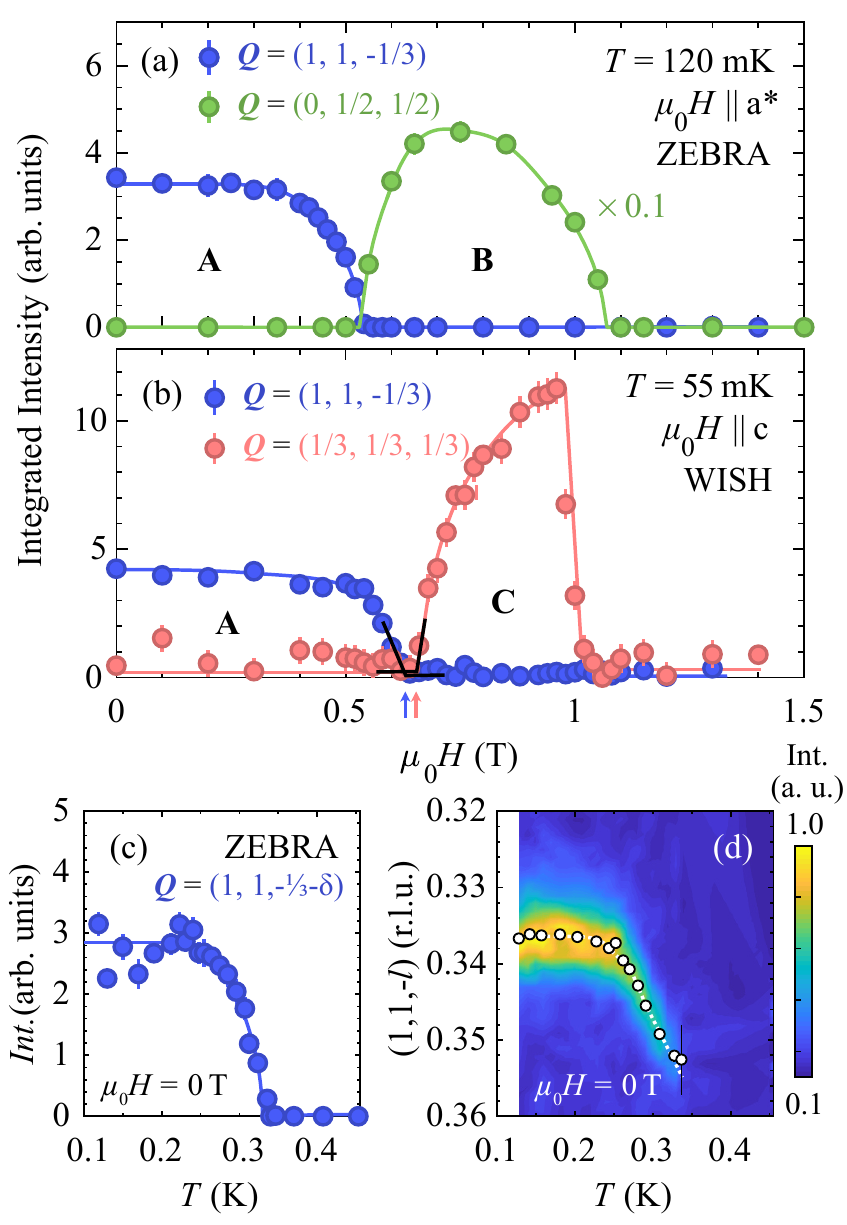}
\caption{Results from single crystal magnetic neutron diffraction.  (a,b) Field dependence of the integrated neutron intensity at the magnetic propagation vectors for: (a) $\textbf{H}\parallel\textit{a}^*$ (ZEBRA, PSI) and (b) $\textbf{H}\parallel\textit{c}$ (WISH, ISIS).  Note that in (a) the intensity of the (0,1/2,1/2) reflection has been rescaled by $\times$0.1.  In (b) the limits of the ordered phases are highlighted and shown with arrows. (c) Evolution of the integrated intensity of the reflection (1,1,-1/3) with temperature at zero magnetic field.  (d) Incommensuration of the propagation vector at zero field against temperature, shown as a shift in the peak position of the (1,1,$l$) reflection.}
\label{fig:Zebra}
\end{figure}

We resorted to single-crystal neutron diffraction to investigate the magnetic structures realized in the low field and the plateau phases.  Figure~\ref{fig:Zebra} summarizes the results obtained from the different instruments. The field dependence of the order parameter is depicted for both field configurations,  which is in perfect agreement with our thermodynamic measurement data. 

Zero field data from both experiments unveil a commensurate phase with propagation vector $\textbf{Q}$ = (0, 0,1/3). Fig. ~\ref{fig:Zebra}(a) and Fig. ~\ref{fig:Zebra}(b) show that magnetic reflection (1,1,-1/3) is present throughout phase A for both field orientations.  The phase is consistent with fully commensurate order, which leads to the appearance of a magnetic supercell, as is shown in Fig.~\ref{fig:Crystal_structure}(d).  Integrated intensity of reflection (1,1,-1/3) drops at the intermediate transition field, above which a different type of order is found depending on the direction of the magnetic field.  For phase B ($\textbf{H}\parallel\textit{a}^*$) we found magnetic reflections (0,1/2,1/2) and (1/2,0,1/2).  These reflections vanish at fields slightly below saturation. Finally, phase C ($\textbf{H}\parallel\textit{c}$) has been found to realize order with propagation vector (1/3,1/3,1/3). Magnetic reflection (1/3,1/3,-1/3), which is inequivalent to the former, has also been found.  Fig.~\ref{fig:Zebra}(b) shows an abrupt drop in the intensity of reflection (1/3,1/3,1/3), consistent with a first order transition to saturation.  

An external magnetic field induces a ferromagnetic component in every lattice site that gives rise to the bulk magnetization. This produces extra scattering proportional to the square of the induced magnetic momentum at the position of each nuclear peak . Fig.~\ref{fig:Magnetization} shows the uniform magnetization density extracted from two nuclear reflections: (020) for $\textbf{H}\parallel\textit{a}^*$ and (200) for $\textbf{H}\parallel\textit{c}$.  We selected reflections where nuclear contribution is minimal while a measurable magnetic intensity can be observed.  The zero-field integrated intensity is subtracted from the data in a field to obtain the corresponding magnetic scattering. Longitudinal magnetization is then plotted as the square root of mangetic intensity. The agreement with bulk measurements is remarkable and further highlights the existence of magnetization plateaus regardless of field orientation. 

Finally,  zero field neutron diffraction reveals an incommensurate state between the low temperature ordered phase and the paramagnetic phase. The onset of incommensurate magnetic order appears around 0.34 K at the wavevector $\textbf{Q} = (0,0,1/3+\delta)$.  Temperature dependence of the  intensity around the (1,1,$l$) reflection in Fig.~\ref{fig:Zebra}(d),  where the peak position is superimposed, shows this incommensuration.  Reduction of the temperature leads to a change in the incommensurate propagation vector roughly linearly with temperature. At 0.26 K the propagation vector locks into the commensurate $\textbf{Q} = (0,0,1/3)$, as observed for the low temperature structure.  The robustness of this evolution to commensuration has been verified for several additional magnetic reflections. 

\section{Discussion}

The purported breathing kagome structure is shown in Fig.~\ref{fig:Crystal_structure}.  Unequal Nd-Nd distances and Nd-O-Nd angles result in inequivalent exchange parameters for neighboring corner-sharing triangles \cite{ashtar2020new}.  This is represented by the exchange constants $J_\bigtriangleup$ and $J_\bigtriangledown$, respectively. However,  a crystallographic analysis cannot rule out the existence of interaction between adjacent kagome planes.  Due to the short distances between kagome planes, the topology of the exchange interaction in \NBWO is likely three dimensional. In fact, the shortest superexchange Nd-O-Nd pathway (nearest neighbors, $J_1$) links rare-earth ions belonging to different kagome planes [Fig. ~\ref{fig:Crystal_structure}(b)]. These couplings are arranged into isolated twisted 3-legged \textit{spin tubes}, one-dimensional structures that extend perpendicular to the kagome planes [see Fig.~\ref{fig:Crystal_structure}(c)].  Noteworthy, the resulting structure considering only nearest neighbor coupling is bipartite. A single tube would show no frustration, highlighting the relevance of further neighbor interactions. A three-dimensional structure with several exchange parameters may have to be regarded, as opposed to the originally suggested kagome structure.  Yet, the onset of static magnetic order is extremely suppressed by the strong magnetic frustration $f = -\theta_W/T_N \approx 12.6 $, confirmed from magnetic susceptibility. 

Six magnetic rare-earth Nd$^{3+}$ ions occupy general Wyckoff positions in the unit cell. The reduced point symmetry around the Nd$^{3+}$ ions [Fig. ~\ref{fig:Crystal_structure}(e)] fully lifts the degeneracy of the total angular momentum levels ($J$ = 9/2) into five Kramers doublets. The strong CEF isolates a single Kramers doublet with a large gap to excited multiplets.  The obtained zero-field entropy is consistent with a value of $S = R\ln(2)$. These two observations show that \NBWO can be described as an effective spin $S$ = 1/2 system below 100 K. However,  the low symmetry precludes attempts to identify unequivocally a CEF-Hamiltonian and to extract the eigenstates of the lowest energy multiplet.   

Both magnetization and susceptibility suggest very little magneto-crystalline anisotropy.  Susceptibility measurements suggest no preferential direction in the high temperature paramagnetic state.  In addition, low-temperature magnetization in the fully saturated pseudospin phase shows no increase up to the highest probed fields.  The increase of magnetization may be a rough estimator of the eigenstate admixing due to anisotropies (via Van-Vleck terms). No appreciable change in magnetization is observed up to 2 T, indicating the total magnetization in the restricted pseudospin subspace is likely to be an approximately good quantum number. It is, thus, likely that the low energy physics in \NBWO can be described in terms of a highly symmetric spin Hamiltonian. A small axial anisotropy may be needed to account for the sharp features found for $\textbf{H}\parallel\textit{c}$.

\begin{figure}[tbp]
\includegraphics[scale=1]{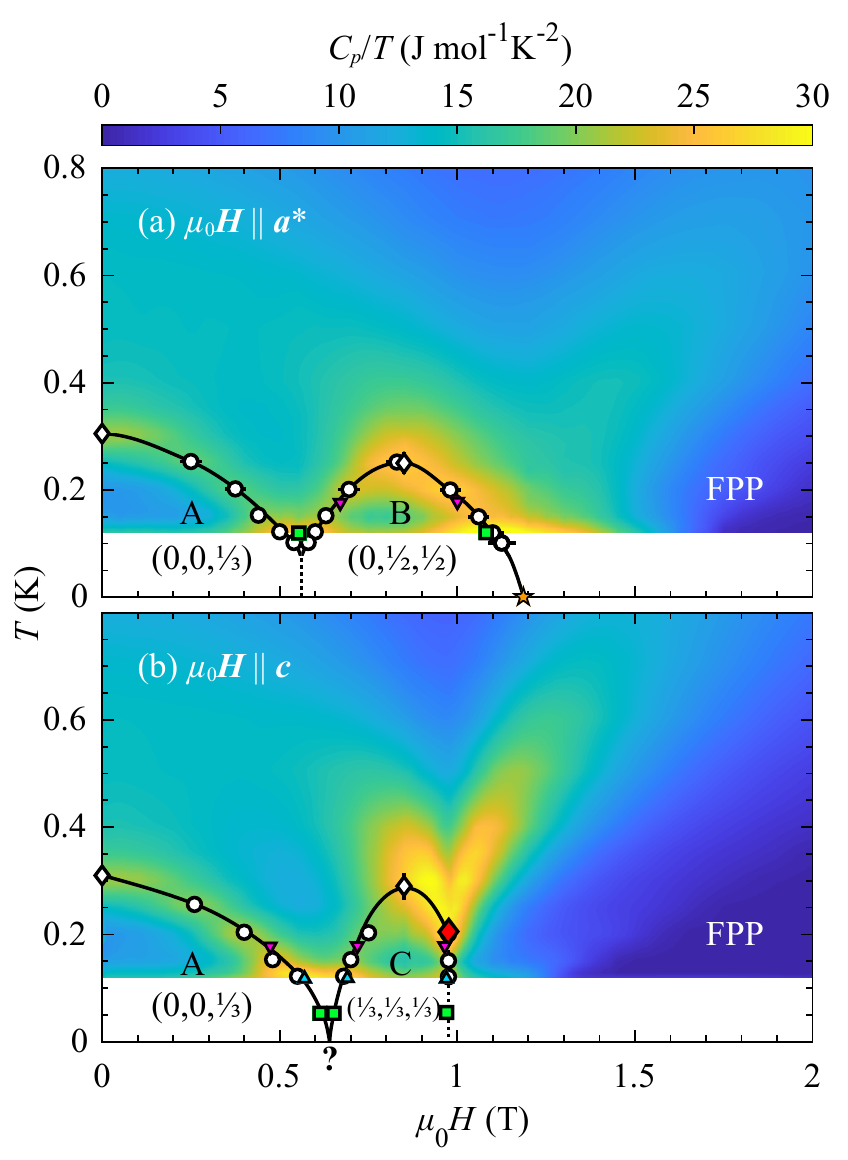}
\caption{Magnetic phase diagram of \NBWO in a magnetic field applied along the principal directions: (a) \textit{a}$^*$ and (b) \textit{c}.  The background depicts false color maps of $C_p (H,T)$,  with a shared color scale.  Symbols: white circles and diamonds represent transitions obtained from field and temperature scans of specific heat, respectively.  Green squares represent the phase boundaries extracted from neutron diffraction data in Fig.~\ref{fig:Zebra}.  Upward-facing blue triangles show transitions extracted from bulk magnetization, downward facing pink triangles transitions from magnetic torque. A red diamond denotes the estimated position of the tricritical point for $\textit{H}\parallel\textit{c}$. An orange star shows the upper critical field estimated in Fig.~\ref{fig:MCE_entropy}. Solid and dashed lines are a guide to the eye, representing second and first order transitions, respectively.  The different phases are labeled as: A, B, C and Fully Polarized Pseudospin (FPP).  The ordered phases show their corresponding magnetic propagation vector, as discussed in the text.}
\label{fig:Phase_diagram}
\end{figure}

To map out the phase diagram in the low temperature regime for \NBWO we use specific heat measurements. Using a combination of all outlined techniques,we identify several regions of magnetic order.  As shown in Fig.~\ref{fig:Phase_diagram}, the system reveals complex behaviour, with two different domes of long-range order observed for each configuration. 

A low field phase (A) extends roughly up to 0.6 T for both studied orientations. This phase possesses commensurate order with propagation vector $\textbf{Q} = $(0,0,1/3). Magnetization measurements show that this phase is hardly magnetizable, suggesting a gapped state in this field range. Although further analysis is needed to understand the magnetic structures of the different phases in detail,  a series of general remarks can be deduced from the data. For phase A, the presence of reflections (0,0,$\pm$2/3) forbids the existence of a collinear structure with spins parallel to \textit{c}. Thus, a coplanar structure in the \textit{ab} plane is likely realized. 

By increasing the magnetic field the system transitions into a field-induced ordered phase.  A field $\textit{H}\parallel\textit{a}^*$ leads to the fractional $m = $ 1/3 plateau phase B, characterized by a propagation vector $\textbf{Q} = $(0,1/2,1/2). The additional presence of wavevectors (1/2,0,1/2) and equivalent suggests a multi-Q structure or the presence of domains in the B phase.  

Strikingly, the order realized in the plateau is completely different when fields are applied in the basal \textit{ab} plane or perpendicular to it. In a field $\textit{H}\parallel\textit{c}$, phase C is found with propagation vector $\textbf{Q} = $(1/3,1/3,1/3).  In contrast, saturation $\textit{H}\parallel\textit{c}$ occurs through a sharp first order phase transition. Magnetocaloric effect supports this claim. A tricritical termination point appears where first and second order transition lines converge as shown in Fig.~\ref{fig:Phase_diagram}(b), at 0.20 K and 0.975 T. The presence of magnetic reflections (1/3,1/3,-1/3) and equivalent also indicates a complex spin texture, with either a multi-Q structure or the presence of domains.While here domains may be consistent with the observed first order transition to saturation, it is not possible at this stage to exclude either possibility.  

The existence of a tricritical point only for one orientation may be related to the large spin-lattice interaction stemming from strong spin-orbit coupling. The transition to saturation for $\textbf{H}\parallel\textit{c}$ can be prematurely precipitated via an 'order by distortion' \cite{tchernyshyov2002order} mechanism. A gain in magnetic energy compensates a small loss in elastic energy, leading to a first order transition to saturation. Though our neutron diffraction data show no evident change in the space group or lattice parameters in the high field phase,  a detailed study would be necessary to discard this possibility. 

Phases A and B appear to merge below 100 mK at 0.56 T. A first order phase transition is speculated between A and B, with a termination bicritical point where all phase boundaries meet. Neutron diffraction data in Fig.~\ref{fig:Zebra}(a) indicate the phases will likely merge slightly below 120 mK.  Interestingly,  between A and C the phase boundaries seem to develop smoothly down to the lowest measured temperatures and converge at $T=0$.  Neutron data at 55 mK show the phases are still separated by paramagnetism at this temperature Fig.~\ref{fig:Zebra}(b).  A highly non-trivial order-to-order quantum phase transition may take place between A and C at zero temperature (indicated with a question mark).  Precise measurements in the vicinity of these phase transitions would provide important insight on their nature.  However, the strong signal from nuclear degrees of freedom and the extremely low temperatures involved prevent further investigation. 

The double hump features in specific above the transition temperature represent a crossover from the low field disordered phase to the high field polarized phase. Such features can be understood in terms of models of hard-core bosons and are usually associated with quantum critical behaviour in one dimensional magnets \cite{korepin1990time, Sachdev1994Finite}.  They can be observed in several quasi-1D antiferromagnets \cite{Blosser2018Quantum,Ruegg2008Thermo}, and therefore suggest the relevance of one-dimensional correlations for the physics of \NBWO. These modulations are accentuated when the field is applied along the direction of the spin tubes ($H\parallel \textit{c}$). Notably, despite the first-order nature of the transition these modulations are still present and seem to be most prominent around the tricritical termination point.  

Plateaux in the magnetically ordered sector are a hallmark of frustrated magnets. The existence of magnetization plateaus (and particularly at 1/3 of saturation) has been predicted for both kagome antiferromagnets \cite{sakai2011critical,capponi2013numerical}, as well as for a model of isolated spin tubes with a weak triangular rung interaction (see Fig.~\ref{fig:Crystal_structure}(c)) \cite{fouet2006frustrated,yonaga2015ground,sakai2012exotic}. The presence of magnetization plateaux independent of the orientation of the applied magnetic field suggests an stabilizing interplay between frustration mechanisms.  

Finally,  we comment on the origin of the observed incommensurate-commensurate (IC-C) transition.  Dipolar interactions are not uncommon in the study of rare-earth based magnets due to their large magnetic moments ($\mu($Nd$^{3+}) = 3.6 \mu_B$) \cite{Palmer2000Order}. Their stabilizing role on incommensurate structures at temperatures above commensurate order has been argued in several systems with hexagonal structure \cite{shiba1982incommensurate,suzuki1983microscopic,gekht1990phase}. The realization of a IC-C transition at zero field opens the question to the importance of dipolar coupling for the low temperature properties of \NBWO. 

We conclude the discussion by comparing \NBWO to its isostructural compounds. To this point, only two other systems in the R$_{3}$BWO$_9$ family have been studied at low temperatures. NMR spectra reveal an inconmmensurate magnetic structure in Sm$_{3}$BWO$_9$\cite{zeng2022incommensurate}, while a dynamical state has been proposed for Pr$_{3}$BWO$_9$ at temperatures as low as 90 mK  \cite{zeng2021local}.  These two systems have been analyzed in terms of 2D Hamiltonians based on the existence of the kagome planes. However, our work highlights the presence of three-dimensional couplings and the potential dominance of the one-dimensional spin tubes.  The discussion outlined here is inevitably relevant for investigations on other members of the family of R$_{3}$BWO$_9$.  

\section{Conclusion}

We have presented a comprehensive study of the low temperature physics of the highly frustrated quantum antiferromagnet \NBWO. Calorimetric and neutron scattering data support the realization of strongly interacting effective spin-1/2 moments below 100 K. Our measurements reveal a complex magnetic phase diagram below 300 mK,  featuring magnetization plateaux for all field orientations.  The ordering brings about important insight about the relevant magnetic interactions.  Different magnetic structures are realized in the plateau states, depending on the direction of the magnetic field. Even though the phase diagram is considerably anisotropic, it can be described in terms of an effective $S$ = 1/2 pseudospin.  

The experimental framework provided here is key for future studies on \NBWO and in the remaining members of the R$_{3}$BWO$_9$.  The presence of the spin-tube structures perpendicular to the kagome planes is indicates that the magnetic properties of these highly frustrated systems cannot be understood in terms of kagome-lattice physics. Further work is needed to fathom the effective dimensionality of the magnetic lattice. 

\section{Acknowledgements}

This work is supported by a MINT grant of the Swiss National Science Foundation.


\bibliography{PRB22_main}


\end{document}